\documentclass[aps,showpacs,twocolumn,superscriptaddress]{revtex4}
\usepackage{graphicx}
\usepackage{amssymb}
\usepackage{epstopdf}
\usepackage{color, amsmath, bm}
\usepackage[hidelinks]{hyperref}
\usepackage{natbib}
\newcommand{\be}{\begin{equation}}
\newcommand{\ee}{  \end{equation}}
\newcommand{\ba}{\begin{eqnarray}}
\newcommand{\ea}{  \end{eqnarray}}
\newcommand{\ket}[1]{\left|#1\right>}
\newcommand{\bra}[1]{\left< #1 \right|}

\RequirePackage{amsmath}
\usepackage{amsfonts}   
\usepackage{physics}

\newcommand{\nb}[1]{\text{n}_{\text{B}}(#1)}
\newcommand{\del}[1]{\delta\left(#1\right)}
\newcommand{\w}[0]{\Tilde{\omega}_0}
\newcommand{\oo}[0]{\Tilde{\Omega}}
\newcommand{\y}{\Tilde{k}_2}
\newcommand{\kk}{\Tilde{k}_1}
\newcommand{\pa}[1]{\left(#1\right)}
\newcommand{\cor}[1]{\left[#1\right]}

\begin{document}

\title{Thermal corrections to quantum friction and decoherence: a Closed-Time-Path approach to atom-surface interaction}

\author{Ludmila Viotti}
\author{M. Bel\'en Far\'\i as}
\author{Paula I. Villar}
\author{Fernando C. Lombardo}
\affiliation{Departamento de F\'\i sica {\it Juan Jos\'e
Giambiagi}, FCEyN UBA and IFIBA CONICET-UBA, Facultad de Ciencias Exactas y Naturales,
Ciudad Universitaria, Pabell\' on I, 1428 Buenos Aires, Argentina.}
\date{\today}                                           

\begin{abstract}  
In this paper we study the dissipative effects and decoherence induced on a particle moving at constant speed in front of a dielectric plate in quantum vacuum, developing a Closed-Time-Path (CTP) integral formulation in order to account for the corrections to these phenomena generated by finite temperatures. We  compute the frictional force of the moving particle and find that it contains two different contributions: a pure quantum term due to quantum fluctuations (even present  at vanishing temperatures) and a temperature dependent component generated by thermal fluctuations (bigger contribution the higher the temperature). We further estimate the decoherence timescale for the internal degree of freedom of the  quantum particle. As expected, decoherence time is reduced by temperature, however, this feature is stronger for large velocities and for resonant situations. When the particle approaches relativistic speed, decoherence time becomes independent of temperature. The finite temperatures corrections to the force or even in the decoherence timescale could be used to track traces of quantum friction through the study of the velocity dependence since the solely evidence of this dependence provides an indirect testimony of the existence of a quantum frictional force.
\end{abstract}

\maketitle

	\section{Introduction}

	One of the most exciting features of modern quantum field theory consists of the nontrivial structure of the vacuum state and the zero-point or vacuum fluctuations \cite{book_milonni}. 
	Among the most remarkable observable consequences of quantum vacuum fluctuations, we can mention the Casimir static force between neutral objects 
	experimentally demonstrated \cite{Lamoreaux1997,Mohideen1998,Ederth2000,Chan2001,Bressi2002,Decca2003}. 
	A less celebrated and renowned phenomena is the appearance of a dissipative force when two neutral lossy bodies are placed at a short distance and set into relative parallel motion at constant speed  \cite{barton_atom_halfspace,intravaia_acceleration,dalvit_intravaia,behunin_hu_atom,volokitin_persson,farias_friction,review_friction,milton2016reviewfriction}. This force is known as quantum friction (QF) and is said to be due to the exchange of Doppler-shifted virtual photons. However, its prediction has inspired a lengthy debate on its origin \cite{pendry97,pendry_debate}.  Due to its short range and small  magnitude, precision measurements  of quantum forces are incredibly difficult and the quantum frictional force has eluded experimental detection so far. Many efforts have been put lately into trying to find conditions that would enhance the force, such as considering non-parallel motion \cite{klatt_farias}, and using promising 2D materials belonging to the graphene family \cite{farias_graphene,farias2018quantum,volokitin2011quantum}. Even though many studies have found some situations for which the force would be increased in several orders of magnitude, its experimental demonstration is still due. Lately, some authors suggested to track traces of quantum friction through the dependence upon the velocity of some other measurable property of the system \cite{klatt2016,farias2019traces}. 
		
	Frictional and normal Casimir forces are not the only effects of  vacuum quantum fluctuations.  For any quantum system, the influence of the environment plays a role at a fundamental level: the system's dynamics can no longer be described in terms of  pure quantum states and unitary evolution. From a practical point of view, all real systems interact with an environment to a greater or lesser extend, which means that we expect their quantum evolution to be altered by decoherence. In the particular case of vacuum fluctuations it is important to note that vacuum field is an environment that cannot be switched off: all matter will unavoidably interact with the electromagnetic vacuum. In that fashion, some of us have investigated the possibility of detecting quantum friction through the decoherence of the internal degree of freedom of a particle that moves in front of an imperfect plate \cite{farias2016}, finding that velocity-dependent corrections to the decoherence time can be relevant for certain choices of the material and the particle's polarizability. Traces of quantum friction in the decoherence timescale could, under some circumstances, be easier to detect than the frictional force itself. The loss of coherence of the particle's dipolar moment becomes relevant in any Ramsey interferometry experiment, where the depolarization of the atom could be macroscopically observed by means of the Ramsey fringes. In the case of a Rydberg atom, this phenomenon could be also observed as a decay of the Raby oscillations \cite{ramsey,rabi}.
	 In our present study, we examine the thermal corrections to the frictional effect on the one side; and on the other, we focus on
	how  decoherence time's dependance upon velocity is modified by the environmental temperature. It is important to note that this further consideration of the thermal corrections represents a more real scenario for experimental purposes than previous analysis at $T=0$ done in \cite{farias2016}.	

	This article is a further contribution and extension of previous works by ourselves and collaborators. We shall consider a particle moving in front of a dielectric plate and thoroughly study the decoherence process of the particle's internal degree of freedom. Herein we shall deal with a more realistic scenario in $3+1$ dimensions and account for thermal corrections in the non-relativistic regime. Not only shall we compute the decoherence time but the frictional force as well,  and study the velocity-dependent corrections in this framework. It is important to stress that the main approach used along this manuscript is the development of a Closed-Time-Path (CTP) integral formulation \cite{CTP_schwinger,CTP_keldysh}. This shall be done with a dual purpose: (i) calculate a general expression for the frictional force of the moving particle and (ii) as a tool to evaluate the decoherence time of the internal degree of freedom of the quantum particle in interaction with the vacuum field and the dielectric mirror. The CTP-method has been used in quantum field theory as an approach to non-equilibrium descriptions of dynamical problems, where dissipative effects arise at the macroscopic level after coarse-graining the detailed information in one or more subsystems, by tracing out those degrees of freedom. In fact, this method presents a combination of both quantum field theory and statistical mechanics.

This article is organized as follows. In the Sec. II, we present the microscopic 
model and use the CTP - approach to functionally evaluate the in-in generating functional. In Sec. III, we evaluate the thermal corrections to the quantum frictional force and analyze the force dependence with the velocity at different temperatures 
of the fields. In Sec. IV we further calculate the influence functional which allows us to estimate decoherence times in Sec. V. Finally, we include a Conclusions Section.

\section{The system}\label{system_section}
\begin{figure}[h]
\centering
\includegraphics[width=\columnwidth]{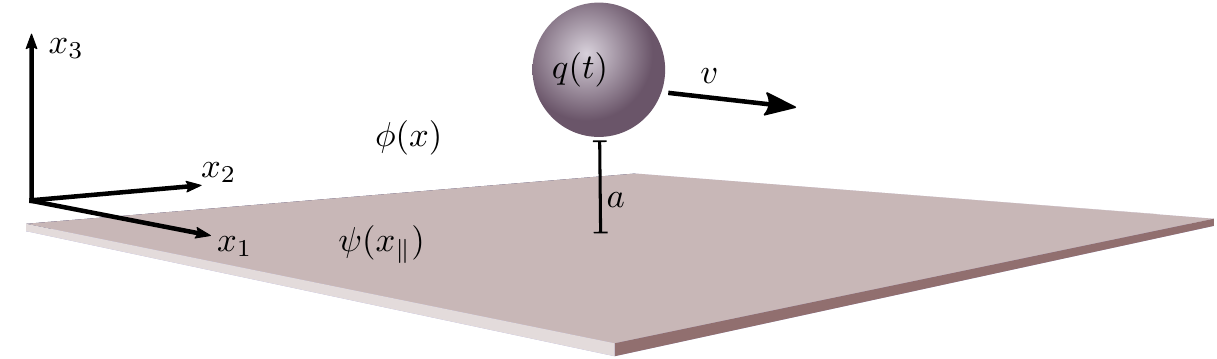}
\caption{\label{esquema} A scheme of the system under consideration, where $\phi(x)$ is the vacuum field, $\psi(x)$ are the
internal degrees of freedom of the plate, and q(t) is the internal degree of freedom of the particle, which follows a macroscopic
trajectory in the $x_1$ direction.}
\end{figure}

We shall consider a neutral particle coupled to a vacuum field, whose center of mass traverses with a velocity $v$ relative to the dielectric plate as shown in Fig.\ref{esquema}. The particle moves in a macroscopic, externally-fixed, uni-dimensional trajectory, in a plane parallel to the plate. The distance $a$ between the particle and the plate is also kept constant by an external source. 
The vacuum field is consider to be a non-massive real scalar field $\phi(x)$ that interacts with the internal degrees of freedom of the plate $\psi(x)$.  We  call $x_1$ the direction of movement of the particle, and $x_3$ the direction perpendicular to the plate. We also consider the particle with an internal degree of freedom named $q$, which interacts with the vacuum field.

We may write the classical action for the system as
\begin{align}\nonumber
    S[\phi,\psi,q]=& \, S_0^{\text{vac}}[\phi]+S_0^{\text{pl}}[\psi]+S_0^{\text{part}}[q] \\
    &+S_{\text{int}}^{\text{pl}}[\phi,\psi]+S_{\text{int}}^{\text{part}}[\phi,q] \, .
    \label{accion}
\end{align}
The first three terms on the right-hand side of \ref{accion} are the corresponding action of:  the plate, the particle and the vacuum field, respectively. The last two terms contain the field-plate and the field-particle interactions. Neglecting boundary terms, the Klein-Gordon action for the vacuum field is given by 
\begin{equation}
    S_0^{\text{vac}}[\phi]= - \frac{1}{2}\int dx \phi(x)[\partial^{\mu}\partial_{\mu}-i\epsilon]\phi(x) \, .
    \label{accion_kg}
\end{equation}
The internal degree of freedom of the particle interacts with the vacuum field trough a current  $J(x)$ with contains both the information about the position and trajectory of the particle, and the strength of the coupling. This interaction term is, then,
\begin{equation}
    S_{\text{int}}^{\text{part}}[\phi,q]=\int dx \phi(x)J(x) \, .
    \label{interaccion_particula}
\end{equation}
We use, for the treatment of this problem, the CTP method \cite{CTP_schwinger} and the Feynman-Vernon influence functional (IF) \cite{feynman_vernon}. While working on the CTP or in-in formalism, the different terms of the action must be integrated along a temporal path that is suitable for real-time evaluations (the Schwinger-Keldysh contour).

We wish to consider finite temperatures, particularly the thermal equilibrium scenario in which all the subsystems (vacuum field, particle and material) are at the same inverse temperature $\beta$. In order to consider such thermal initial states, the integration path must be changed to the Kadanoff-Baym contour $\mathcal{C}$, which differs from Schwinger-Keldysh in having an extra branch along the imaginary axis which goes from $-\tau-i\epsilon$ to $-\tau-i\beta$ \cite{KB}.
When integrating over the different fields, one must also evaluate them along it.
A single time representation can be achieved by doubling the degrees of freedom of the system.
If the sources and fields are assumed  to be the configurations $J_+(x),\phi_+(x)$ on the first branch and $J_-(x),\phi_-(x)$ on the second branch;  by considering them as independent fields, we can then write the in-in generating functional as a functional integral over fields along a single time interval, where the boundary condition $\phi_+|_\tau=\phi_-|_\tau=\phi_{\text{out}}(\tau,x)$ implies that the integrals can't be done independently. This representation enables us to write the integrand in the familiar way
\begin{equation}
    Z_{\text{in-in}}[\mathbf{J}]=\oint\mathcal{D}\phi e^{-i\int dx \pa{\mathbf{\phi}^T\frac{\hat{K}}{2}\mathbf{\phi}-\mathbf{\phi}\mathbf{J}}},
    \label{in_in_generating_functional}
\end{equation}
where now, in contrast with the usual in-out generating functional, the differential operator $\hat{K}$ is a $2\times2$ matrix. 
The integration over the vacuum field $\phi$ can be easily done, resulting in the known expression, 
\begin{equation}
    Z_{\text{in-in}}[\mathbf{J}]= e^{-\frac{1}{2}\int dx dx' J_{\alpha}(x)G_{\alpha\beta}(x,x')J_{\beta}(x') },
\end{equation}
where a summation is assumed over repeated indexes, and the free field propagator $G_{\alpha \beta}(x,x')$ is a $2\times 2$ matrix.  In this case, the free propagator for a massless scalar field is given in momentum space by
\begin{align}\nonumber
    G^{(0)}(k)=&\left(\begin{array}{cc}
         \frac{1}{k^2+i\epsilon}&-2\pi i \delta(k^2)\theta(-k^0)  \\
         -2\pi i \delta(k^2)\theta(k^0) &\frac{-1}{k^2-i\epsilon} 
    \end{array}\right)\\
    &-2\pi i \text{n}_B(|k^0|)\delta(k^2)
    \left(\begin{array}{cc}
        1 &1  \\
        1  &1 
    \end{array}\right) ,
\end{align}
where $\text{n}_B$ is the Bose distribution ${\text{n}_B(|k^0|)=\frac{1}{e^{|k^0|/T}-1}}$, with $T=1/ \beta$. We are working with units such that $k_B=\hbar=1$.
We aim to obtain the in-in effective action of the system $\Gamma$, which is defined as
\begin{equation}
    e^{i\Gamma}=\int_{\mathcal{C}} \mathcal{D}\phi\mathcal{D}\psi\mathcal{D}q\hspace{.2cm}e^{iS[\phi,\psi,q]}.
    \label{accion_afectiva_sistema}
\end{equation}
Rather than integrating over every degree of freedom  simultaneously, it is convenient to introduce a partial result by integrating out the degrees of freedom of the plate and the particle so that the whole information about them is contained in an effective interaction term of the form $S_{\text{int}}^{\text{vac}}[\phi]=S_{\text{pl}}^{\text{vac}}+S_{\text{part}}^{\text{vac}}$. As we are assuming locality in the microscopic interactions, we know $S_{\text{pl},\text{part}}^{\text{vac}}$ depends on $\phi(x)$ only for $x^{\mu}$ inside the region defining the respective element. Thus, under the assumption that, either exactly (as in the model we consider) or approximately, $S_{\text{pl},\text{part}}^{\text{vac}}$ is quadratic, we have \cite{fosco2011_sidewise}
\begin{equation}
    S_{\text{int}}^{\text{vac}}[\phi]=-\frac{1}{2}\int dxdx' \phi(x)V(x,x')\phi(x'),
\end{equation}
with 
\begin{equation}
    V(x,x')=V^{\text{part}}(x,x')+V^{\text{pl}}(x,x'),
\end{equation}
and
\begin{align}
    V^{\text{part}}(x,x')=&\delta\pa{x_3-a}\delta\pa{x_2}\delta\pa{x_1-vx_0}\\ \nonumber
    &\times\delta\pa{\mathbf{x}-\mathbf{x'}}g(x_0-x_0')\\
    V^{\text{pl}}(x,x')=&\delta\pa{x_3}\delta\pa{x_3-x_3'}\lambda(x_{\parallel}-x_{\parallel}'),
\end{align}
where the functions $g(x_0)$ and $\lambda(x_\parallel)$ depend on the microscopic models, and $x_{\parallel}$ refers to the coordinates which are parallel to the plate $(x_0,x_1,x_2)$.
These effective potentials can be determined by considering a specific microscopic model, or even introduced {\textit ad-hoc} based on particular assumptions.
In any case, the integral over $\phi$ becomes a Gaussian 
\begin{equation}
\label{funcional_generatriz_fi}
    \mathcal{Z}_{\text{in-in}}=\int \mathcal{D}\phi\hspace{.2cm}e^{-\frac{i}{2}\int dx dx' \phi(x)\Hat{K}\phi(x')}.
\end{equation}
The differential operator $\Hat{K}$ appearing on Eq. \eqref{funcional_generatriz_fi} is
\begin{equation*}
    \Hat{K}=\Hat{K}_0(x,x')\del{x-x'}-V^{\text{pl}}(x,x')-V^{\text{part}}(x,x'),
\end{equation*}
where $\Hat{K}_0$ is the differential operator for a free field without considering neither the plate nor the particle. We want to find $\Hat{K}^{-1}$ such that $\Hat{K}\Hat{K}^{-1}=\del{x-x'}$.
Considering this challenge, we notice that the effective potentials $V(x,x')$ are proportional to some coupling constants $\lambda$ and $g$ between the vacuum field and the internal degrees of freedom of the plate and particle.  If these couplings are weak, we can obtain a perturbative expression in $\lambda$ and $g$ for $\Hat{K}^{-1}$
\begin{equation}
    \label{expansion_propagador}
    G_{\alpha\beta}=G_{\alpha\beta}^{(0)}+G_{\alpha\gamma}^{(0)}V^{\text{pl}}_{ \gamma\delta}G_{\delta\mu}^{(0)}V^{\text{part}}_{\mu\nu}G_{\nu \beta}^{(0)} + \text{part}\leftrightarrow\text{pl}.
\end{equation}
The integrals involved in the contractions were omitted to simplify the notation.
By functionally integrating \eqref{funcional_generatriz_fi} over $\phi$ we obtain the effective action for the whole system.

\subsection{Microscopic model}\label{model}

In this section we  must characterize the system under study. We have defined so far a real massless scalar field $\phi$
which interacts with another field $\psi$, describing  the internal degrees of freedom of a plate (but  in fact could be  associated to any other system that we might be interested in).  When we integrate those degrees of freedom out, we obtain a non-local effective potential $V(x,x')$ that contains the information about the characteristics of the plate and particle.
Hence we consider a thin infinitesimal plate occupying the $x_3=0$ plane, with internal degrees of freedom that behave as one-dimensional harmonic oscillators, one at each point of the mirror. They have generalized coordinates $Q(x_{\parallel})$ taking values in an internal space.  No coupling between the oscillators is included. We do consider a linear coupling between each oscillator and the vacuum field. The terms in the system action depending on $Q$,
\begin{equation*}
    S_0^{\text{pl}}=\frac{1}{2}\int dx\delta(x_3)\left[\Dot{Q}(x_{\parallel})-(\Omega^2-i\epsilon)Q^2(x_{\parallel})\right]
\end{equation*}
and
\begin{equation*}
    S_{\text{int}}^{\text{pl}}=q\int dx \delta(x_3)Q(x_{\parallel})\phi(x),
\end{equation*}
which results in a Gaussian functional integral to find $S_{\text{pl}}^{\text{vac}}[\phi]$ and therefore the effective potential $V^{\text{pl}}$, which will be a $2\times2$ matrix given, in momentum space, by \cite{fosco2011_sidewise}
\begin{align}\label{placa}
    V^{\text{pl}}_{\alpha\beta}(k_{\parallel},k_{\parallel}',x_3,x_3')=&(2\pi)^3\delta^{(3)}(k_{\parallel}-k_{\parallel}')\lambda_{\alpha\beta}(k^0)\\
    &\times \delta(x_3-x_3')\delta(x_3) \nonumber.
\end{align}    
The particle is considered to be punctual, moving along the $x_1$ axis with a constant velocity $v$ at a fixed distance $x_3=a$ above of the plate, and interacting locally in position with the vacuum field, then the current $J(x)$ in Eq. (\ref{interaccion_particula}) will have the form
\begin{align}
\label{eq:J}
J(x)=g \, q(x_0)\delta\pa{x_1-vx_0}\delta(x_2)\delta(x_3-a) \, ,
\end{align}
where $g$ is the coupling constant between the vacuum and the internal degree of freedom of the particle. Therefore, the effective potential corresponding to the particle becomes \cite{fosco2011_sidewise}

\begin{widetext}
	\begin{align}\nonumber
	V^{\text{part}}_{\alpha\beta}(k_{\parallel},k_{\parallel}',x_3,x_3')=2\pi\delta(x_3-x_3')\delta(x_3-a)
	g_{\alpha\beta}(k^0-vk^1)\delta\left(k^0-vk^1 -(k'^0-vk'^1) \right),
	\end{align}
	with
\begin{equation}
    \lambda_{\alpha\beta}(k^0)=\lambda^2\left(\begin{array}{cc}
         \frac{1}{(k^0)^2-\Omega^2+i\epsilon}&- \frac{\pi }{\Omega} i\delta(k^0+\Omega) \\
         - \frac{\pi }{\Omega} i\delta(k^0-\Omega)&\frac{-1}{(k^0)^2-\Omega^2-i\epsilon} 
    \end{array}\right)-2\pi i\lambda^2 \text{n}_B(|k^0|)\delta((k^0)^2-\Omega^2)
    \left(\begin{array}{cc}
        1 &1  \\
        1  &1 
    \end{array}\right) \, ,
\end{equation}
and
\begin{equation}
    g(k^0)=g^2\left(\begin{array}{cc}
         \frac{1}{(k^0)^2-\omega_0^2+i\epsilon}&- \frac{\pi }{\omega_0} i\delta(k^0+\omega_0) \\
         - \frac{\pi }{\omega_0} i\delta(k^0-\omega_0)&\frac{-1}{(k^0)^2-\omega_0^2-i\epsilon} 
    \end{array}\right)-2\pi ig^2 \text{n}_B(|k^0|)\delta((k^0)^2-\omega_0^2)
    \left(\begin{array}{cc}
        1 &1  \\
        1  &1 
    \end{array}\right) \, .
\end{equation}

Here $\omega_0$ is some characteristic frequency of the particle internal degree of freedom $q$, and $\Omega$ is the characteristic frequency of the harmonic oscillators constituting the plate. 
\end{widetext}
The case of a mirror imposing `perfect", i.e., Dirichlet boundary conditions can be obtained by taking particular limits in the definition of Eq.(\ref{placa}). This Dirichlet limit may be reached by assuming $\lambda^2/\Omega^2  \rightarrow \infty$ in the propagators (it would be similar for Neumann boundary conditions). It is easy to check that 
$V^{\text{pl}}_{\alpha\beta}(k_{\parallel},k_{\parallel}',x_3,x_3')= {\tilde\lambda}^2 \delta^{(3)}(k_{\parallel}-k_{\parallel}')\delta(x_3-x_3')\delta(x_3)$ with $ {\tilde\lambda}^2\rightarrow \infty$, implies Dirichlet boundary conditions. As it could be seen below, the quantum frictional force is zero in the Dirichlet limit. The microscopic model for the material 
is a toy model in which absorption is neglected.


\section{Finite temperature frictional force}\label{friction_force}

When the particle and the plate are in relative motion,
there is an energy transfer to the system \cite{farias2016}.
Therefore, energy conservation implies that there should be
some force performing mechanical work when moving the
particle.
Moreover, since this motion has a constant speed,
the force has to be dissipative in nature. However, the system is not completely
closed  because the
particle is forced to move with constant speed and is moreover kept at
a fixed height for external agents (that are not further considered
in the calculations).

\begin{figure}[h]
\centering
\includegraphics[width=\columnwidth]{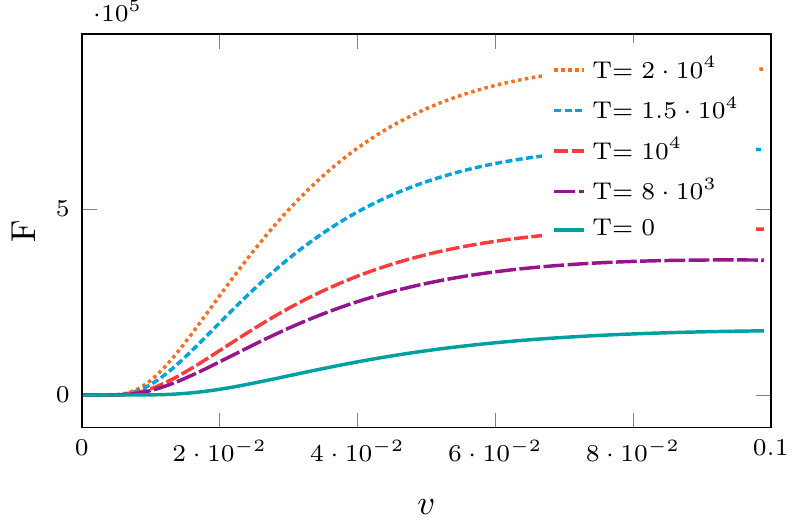}
\caption{\label{grafico_friccion} Dissipative force, as a function of the relative velocity $v$ between the plate and the particle, for $\w=0.03$, $\oo = 0.01$ and $a=10^{-6}~[m]$ and $T$ as defined in Sec. \ref{system_section}.}
\end{figure}

In order to find an expression for that force, we compute
the mean value of the energy-momentum tensor $t_{\mu\nu}$ in vacuum and in the steady regime $\expval{t_{\mu\nu}}=\bra{0_{\text{in}}}t_{\mu\nu}\ket{0_{\text{in}}}$. The frictional force between the particle and the plate can be obtained by means of the point-splitting technique as 
\begin{equation}
    F=\lim_{x\rightarrow a^+}\expval{t_{13}(x)}-\lim_{x\rightarrow a^+}\expval{t_{13}(x)},
    \label{point_splitting}
\end{equation}
where
\begin{align}
    \expval{t_{13}(x)}&=\lim_{x'\rightarrow x}\expval{\partial_1\phi(x)\partial'_3\phi(x')}\\
    &=\frac{1}{2}\lim_{x'\rightarrow x}\int\frac{dp^0}{2\pi}\frac{d^2p_{\parallel}}{(2\pi)^2}(ip_1)\partial'_3G_1(p^0,p_{\parallel},x_3,x'_3). \nonumber
\end{align}
Here, ${G_1(x,x')=\bra{0_{\text{in}}}\phi(x)\phi(x')\ket{0_{\text{in}}}}$ is the Hadamard's two-point function,
which is related to Feynman propagator ${G_{++}(x,x')=\bra{0_{\text{in}}}T\phi(x)\phi(x')\ket{0_{\text{in}}}}$ by $G_1(x,x')=$ $2\Im\pa{ G_{++}(x,x')}$.
Expanding the Feynman propagator $ G_{++}(x,x')$ as in Eq. (\ref{expansion_propagador}),
it is possible to compute every contraction exactly,
thus obtaining a perturbative expression for the desired component of the energy-momentum tensor. 
It is clear from Eq. (\ref{point_splitting}) that the dissipative force is given by the discontinuity of
$\expval{t_{13}(x)}$ at $x=a$. The derivatives can be easily calculated by writing the different terms of the effective propagator, corresponding to different orders in the coupling constants, in momentum space. As the free propagator is continuous at $x=a$, it does not contribute to the force.
It can also be shown that the only non-vanishing contribution to the force comes from those terms of Eq. (\ref{expansion_propagador}) with $\nu=\beta=+$, since the terms with $\nu = -$ are continuous at $x_3=a$. The force is then given by
\begin{align}\nonumber
    F&=\lim_{x'\rightarrow x} \Im 
    \int dv~du~ dz~ dy ~\partial_1G^{(0)}_{+\alpha}(x,y)V^{\rm pl}_{\alpha\beta}(y,z)\\\nonumber
    &\times G_{\beta\gamma}^{(0)}(z,u)V_{part,\gamma+}(u,v)\left[\lim_{x_3'\rightarrow a^+} \partial_3'G_{+ +}^{(0)}(v,x')\right.\\
    & \left.- \lim_{x_3'\rightarrow a^-} \partial_3'G_{+ +}^{(0)}(v,x') \right] + \text{part}\leftrightarrow\text{pl} \, .
    \label{fuerza_feynman}
\end{align}
By Fourier transforming the Klein Gordon propagators and the potentials in the parallel coordinates, both the derivative $\partial_{3'}$ and the limit for $x_3 \rightarrow a$ can be explicitly computed. When doing so, the terms in Eq. (\ref{fuerza_feynman}) with $x_3=0$ vanishes. This procedure adds a factor of (-1) on the remaining term, resulting in a simplified expression for the force 
\begin{align}\nonumber
    F=&-\lim_{x'\rightarrow x} \Im 
    \int \frac{dk_{\parallel}}{(2\pi)^3}(-ik^1)
    G^{(0)}_{+\alpha}(k_{\parallel},a,0)
    \\ &\times \lambda_{ \alpha\beta}(k^0)
    G_{\beta\gamma}^{(0)}(k_{\parallel},0,a)g_{\gamma+}(k^0-vk^1) \, .
    \label{force_momentum_limit}
\end{align}
The integrand consists in eight combinations of the $\alpha, \beta$ and $\gamma$ indexes. These possible combinations lead to a total of 64 terms, most of which vanish due to parity considerations, or as a result of the Heaviside and Dirac delta functions appearing in the propagators and potentials.
The whole contribution to the force can be seen to come from only eleven of those terms.
The combination $\{\alpha, \beta,\gamma\}=\{+--\}$ is completely non-vanishing, while the combinations $\{\alpha, \beta,\gamma\}=\{++-\},\{-+-\}$ and $\{---\}$ have one non-vanishing term each.
\begin{widetext}
Replacing the propagators and functions $g$ and $\lambda$ with their explicit expressions, making use of the Sokhotski–Plemelj theorem \cite{weinberg} and the properties of the delta function, the frictional force is found to be 
\begin{equation}
    F   =-a^2\lambda^2g^2\frac{\w-\oo}{4\w\oo}F_1 -\frac{a^2\lambda^2g^2}{4\w\sqrt{1-v^2}}F_2
    -\frac{a^2\lambda^2g^2}{16\w\oo}F_3
    \label{fuerza}
\end{equation}
with
\begin{align}
    F_1=&\int\frac{d\y}{2\pi}\theta \left[\zeta_-\pa{k^2}\right]\left\lbrace \cos^2\left(\frac{1}{v}\sqrt{\zeta_-\pa{k^2}}\right)\left[\frac{4\nb{\oo}^3+\nb{\oo}^2}{\vartheta} +\frac{\nb{\oo}\pa{4\nb{\oo}+2}-\nb{\w}}{\sqrt{\zeta_-\pa{k^2}}\sqrt{\vartheta}}\right.\right]\\ \nonumber
    &\hspace{2.9cm}-\cos\left(\frac{2}{v}\sqrt{\zeta_-\pa{k^2}}\right)\frac{1}{4}\frac{\nb{\oo}-\nb{\w}}{\zeta_-\pa{k^2}}\Bigg\rbrace,\\
    F_2=&\int \frac{d\y}{2\pi}\kk d\kk \nb{|\kk v-\w|} \, \text{p.v.}\left(\frac{1}{(\kk v-\w)^2-\oo^2}\right)\theta\cor{\zeta\pa{k^1,k^2})}\frac{\sin(2\sqrt{\zeta\pa{k^1,k^2}})}{\sqrt{\zeta\pa{k^1}}\sqrt{\zeta\pa{k^1,k^2}}}~~~{\mathrm {and}}\\    %
    F_3=&- \int \frac{d\y}{2\pi}\left[(\w-\oo)\frac{\nb{\oo}-\nb{\w}}{\zeta_-\pa{k^2}}e^{\frac{-2}{v}\sqrt{-\zeta_-\pa{k^2} }}\theta\cor{-\zeta_-\pa{k^2}}+(\w+\oo) \frac{\nb{\oo}+\nb{\w}+1}{\zeta_+\pa{k^2}}e^{-\frac{2}{v}\sqrt{-\zeta_+\pa{k^2}}}\right]
\end{align}
\end{widetext}where $\zeta\pa{k^1,k^2}= (\kk v-\w)^2-\kk^2-\y^2$,
$\zeta\pa{k^1}= v^2\pa{(\kk v-\w)^2-\kk^2}+\w^2$,
$\zeta_{\pm}\pa{k^2}= v^2\oo^2-(\w\pm\oo)^2-v^2\y^2$ and  $\vartheta=v^2\oo^2-(\w-\oo)^2$.
In order to write Eq. (\ref{fuerza}), we have defined the dimensionless variables $\kk=ak^1$ and $\y=ak^2$, and parameters $\oo=a\Omega$, $\w=a\omega_0$. Thus,  the dimensional dependence to be concentrated in the factor $a^2\lambda^2g^2$. Being the dimensions of each constant $[\lambda]= m^{3/2}$, $[g]=m^{1/2}$ and $[a]=m^{-1}$, the force is found to have the right dimensions. The parameter $\beta$ (the inverse of the temperature) was redefined as $ \frac{\beta}{a}\rightarrow \beta$, rendering it dimensionless as well. 
The theta function $ \theta\left(v^2\oo-(\w-\oo)^2-v^2\y^2\right)$ that appears on $F_1$
 enforces the whole term to vanish unless
\begin{align}
\oo > \frac{\w-v\sqrt{\w^2-(1-v^2)\y^2}}{1-v^2}	\, , \\
\oo < \frac{\w+v\sqrt{\w^2-(1-v^2)\y^2}}{1-v^2} \, .
\end{align}
Considering the extreme situations, these conditions impose the restriction $ \frac{\w}{1+v}<\oo < \frac{\w}{1-v}$ which, in the limit of $v\ll 1$, is rarely satisfied.The behavior of the force will then be given, for non-relativistic motion of the particle, by
\begin{align}
    F&\approx -\frac{a^2\lambda^2g^2}{4\w\sqrt{1-v^2}}F_2
    -\frac{a^2\lambda^2g^2}{16\w\oo}F_3 \, .
    \label{fuerza_vpeque}
\end{align}
In Fig.\ref{grafico_friccion} we show the dissipative (frictional) force on the particle as a function of the 
dimensionless velocity $v$. It can be seen that the force develops two different contributions: one contribution is purely quantum, is generated by the quantum fluctuations of the electromagnetic vacuum, and is present even at vanishing temperature. The other contribution, produced by thermal fluctuations, grows with temperature and is dominant at high temperatures.
For small but non-vanishing temperatures, however, both contributions are of the same order, and a detection of a frictional force at relatively high velocities and small finite temperatures would imply both a detection of a thermal dissipative force and a quantum frictional force. For very small velocities both components vanish, since the energy supplied to the system by moving the particle is not sufficient to excite the internal degrees of freedom of the material (which requires a finite amount of energy determined by the characteristic frequency $\Omega$) and be thus dissipated \cite{fosco2010inertial}.
For velocities slightly higher than this threshold, is the thermal component the one that grows more rapidly, which is 
not surprising since the quantum frictional force, at this order in perturbation theory, is vanishingly small for small velocities \cite{farias_friction,intravaia_acceleration}. However, it is worth remarking that at larger velocities there is definitively a purely quantum contribution, present even at vanishing temperatures, but that thermal fluctuations enhance the total frictional force and become the main contribution for higher temperatures.

From an expansion in powers of the velocity $v$, it is possible to find that
ours results for the friction force predict a linear velocity dependence. 

\section{Influence functional} \label{imfluence_functional_section}
Herein, we shall obtain the generating functional of the particle by integrating out those degrees of freedom corresponding to the plate and the vacuum field in Eq.(\ref{accion_afectiva_sistema})

\begin{equation}
    \mathcal{Z}[q]=e^{i\pa{S_0^{\text{part}}[q]+S_{\text{IF}}[q]}}.
    \label{funcional_q}
\end{equation}

The whole information about the effect of the environment on the particle is contained in $S_{\text{IF}}[q_+,q_-]$. The dynamics of the system (particle) under the influence of the environment (vacuum field and plate) is described by the Feynman-Vernon IF $\mathcal{F}$ which is defined by
\begin{equation}
   \mathcal{F}[q_+,q_-]= e^{iS_{\text{IF}}[q_+,q_-]}:=\int_{\mathcal{C}} \mathcal{D}\varphi_E\hspace{.2cm}e^{iS[\varphi_E,q]},
   \label{funcional_influencia_definicion}
\end{equation}
where $\varphi_E$ represents any environment variable \cite{feynman_vernon}.
After performing the integral over the plate field $\psi(x)$ we will get an effective action for the vacuum field and, as a consequence, a modified vacuum field propagator. Following the steps performed to get Eq. (\ref{expansion_propagador}) we find
\begin{equation}
    G_{\alpha\beta}=G_{\alpha\beta}^{(0)}+G_{\alpha\gamma}^{(0)}V^{\text{pl}}_{\gamma\delta}G_{\delta\beta}^{(0)}.
    \label{expansion_propagador_soloplaca}
\end{equation}

Explicitly computing (\ref{funcional_q}), the influence action can be found to be
\begin{equation}
    S_{\text{IF}}=-\frac{1}{2}\int dx dx' J_{\alpha}(x)G_{\alpha\beta}(x,x')J_{\beta}(x'),
\end{equation}
where summation is assumed over repeated indexes, and $J_{\alpha}$ are the currents appearing in the coupling between the particle and the field (\ref{interaccion_particula}). Considering the perturbative expansion for the propagator (\ref{expansion_propagador_soloplaca}), the influence action can be expanded as
\begin{equation}
    S_{\text{IF}}[q_+,q_-]= S_{\text{IF}}^{(1)}[q_+,q_-]+S_{\text{IF}}^{(2)}[q_+,q_-],
  \label{eq:expIA}  
\end{equation}\\
where $S_{\text{IF}}^{(1)}$ is obtained considering the free Klein-Gordon propagator $G_{\alpha\beta}^{(0)}$, and $S_{\text{IF}}^{(2)}$ is obtained from the second term in the expansion (\ref{expansion_propagador_soloplaca}), $G_{\alpha\beta}^{(1)}$.
It is useful to define
\begin{align}
    \Delta J(x)=\frac{J_+(x)-J_-(x)}{2}\\
    \Sigma J(x)=\frac{J_+(x)+J_-(x)}{2}.
\end{align}
By writing Eq. (\ref{interaccion_particula}) in terms of $\Delta J$ and $\Sigma J$ and exploiting the definition and properties of the different CPT propagators, one obtains, for each term in the expansion for the influence action \eqref{eq:expIA}, the expression \cite{feynman_vernon},
\begin{widetext}
\begin{align}\nonumber
    S_{\text{IF}}^{(i)}[q_+,q_-]=-\frac{1}{2}\int dx dx'& \Big\{ \Delta J(x)\left[G_{++}(x,x')-G_{--}(x,x')-G_{-+}(x,x')+G_{+-}(x,x')\right]\Delta J(x')\\
    &+2\Delta J(x)\left[G_{++}(x,x')+G_{--}(x,x')+G_{-+}(x,x')+G_{+-}(x,x')\right]\Sigma J(x')\Big\}.
    \label{funcional_influencia_i}
\end{align}
\end{widetext}
By using some further considerations concerning the properties of the in-in propagators, it is possible to define  the noise (diffusion) kernel associated to fluctuations and considered as source of decoherence effects
\begin{align}\nonumber
    N_i(x,x')\equiv &-i\pa{G_{++}(x,x')-G_{--}(x,x')}\\
    =& \, 2\Im G_{++}(x,x') \, ;
\end{align}
 and the dissipation kernel
\begin{align}\nonumber
    D_i(x,x')\equiv& \frac{1}{2}\left[G_{++}(x,x')+G_{--}(x,x') \right. \\ \nonumber
    &\left. +G_{-+}(x,x')+G_{+-}(x,x')\right]\\
    =&2\theta(x_0-x_0')\Re G_{++}(x,x') \, .
\end{align}
Both kernels are real quantities, and the dissipation kernel is explicitly causal \cite{feynman_vernon}. 
Considering these definitions, the terms $i=1,2$ in the perturbative expansion of the influence action become
\begin{align}\nonumber
     S_{\text{IF}}^{(i)}[q_+,q_-]=-&\int dx dx' \left[i\Delta J(x)N_i(x,x')\Delta J(x')\right.\\
    &+\left.2\Delta J(x)D_i(x,x')\Sigma J(x')\right].
    \label{funcional_influencia_nucleos}
\end{align}
The equation of motion for the internal degrees of freedom of the particle can be formally written as
\begin{equation}
    \Ddot{q}(t)+\omega_0^2q(t)+\int dt' D(t,t') q(t')=\xi(t),
    \label{eom}
\end{equation}
where dissipation on the particle is originated in the kernel $D$ and fluctuations are
generated by the stochastic force $\xi(t)$, that must fulfill $\langle \xi(t) \rangle=0$ and 
 $\langle\xi(t)\xi(t')\rangle =N(t,t')$. Fluctuations enter into the equation of motion by means of 
\begin{equation*}
    \int \mathcal{D}\xi P[\xi] e^{-i\int dt  \Delta q(t)\xi(t)}=e^{i\int dtdt' \Delta q(t) N(t,t')\Delta q(t')}.
\end{equation*}
where $P[\xi]$ is a Gaussian probability distribution for $\xi$.

From the expression in Eq.\eqref{eq:J} proposed for the current $J_\alpha(x)$, $\Delta J(x)$ and $\Sigma J(x)$ take the form
\begin{align}
    \Delta J(x)&= g\hspace{.2cm}\Delta q(x_0)\delta\pa{x_1-vx_0}\delta(x_2)\delta(x_3-a)\\
    \Sigma J(x)&=g\hspace{.2cm}\Sigma q(x_0)\delta\pa{x_1-vx_0}\delta(x_2)\delta(x_3-a) \, .
\end{align}
We can thus define, for $N\equiv N_1 + N_2$
\begin{widetext}
\begin{align}\nonumber
    N(t,t')&\equiv g^2\int d\mathbf{x}d\mathbf{x'}\delta\pa{x_1-vt}\delta\pa{x_2}\delta\pa{x_3-a}N(x,x')\delta\pa{x_1'-vt'}\delta\pa{x_2'}\delta\pa{x_3'-a}\\
    &=2g^2\Im\int d\mathbf{x}d\mathbf{x'}\delta\pa{x_1-vt}\delta\pa{x_2}\delta\pa{x_3-a}G_{++}(x,x')\delta\pa{x_1'-vt'}\delta\pa{x_2'}\delta\pa{x_3'-a},
\end{align}
\end{widetext}

where we have used the notation $t=x_0$.
Within our model, we can then write explicit expressions for $N_i$, and $D_i$ ($i=1,2$) as integrals in momentum space:
    \begin{equation}
        N_i(t,t')=2g^2\Im\int \frac{dk}{(2\pi)^4}e^{i\pa{k^0-vk^1}(t-t')}G^{(i-1)}_{++}\, ,
        \label{noise}
    \end{equation}
    \begin{equation}
        D_i(t,t')=2g^2\Re\theta(t-t')\int \frac{dk}{(2\pi)^4}e^{i\pa{k^0-vk^1}(t-t')}G^{(i-1)}_{++}\, .
    \end{equation}

\section{Decoherence of the particle's internal degrees of freedom}

Within the consistent histories approach to quantum mechanics, quantum evolution can be considered as a coherent superposition of fine-grained histories. If one defines the c-number $q(t)$ as specifying a fine-grained history, the quantum amplitude for that history is $\psi[q(t)]$ $\sim e^{iS[q(t)]}$ \cite{lombardo2007decoherence}. In the quantum open system approach that we have adopted here, we are concerned with coarse-grained histories
\begin{equation}
    \Psi[\alpha]=\int \mathcal{D}q e^{iS[q(t)]}\alpha[q(t)],
\end{equation}
where $\alpha[q]$ is the filter function that defines the coarse-graining.  At first instance this filtering corresponds to tracing over all the degrees of freedom of the composite environment. From this, we define the decoherence functional for two coarse-grained histories as
\begin{equation}
    \mathcal{D}[q_+,q_-]=\int \mathcal{D}q_+\mathcal{D}q_- e^{i\pa{S[q_+]-S[q_-]}}\alpha_+[q_+]\alpha_-[q_-].
\end{equation}
Decoherence means physically that the different coarse-graining histories making up the full quantum evolution acquire individual reality, and may therefore be assigned definite probabilities in the classical sense. A necessary and sufficient condition for the validity of the sum rules of probability theory (i.e. no quantum interference terms) is
$\Re   \mathcal{D}[q_+,q_-] \approx 0$ 
when $\alpha_+ \neq \alpha_-$ \cite{PhysRevD.46.1580}. Such histories, which can be assigned probabilities consistently as a result of the absence of interference, are consistent histories.
For our particular application, we wish to consider as a single coarse-grained history all those fine-grained ones where the solution $q(t)$ remains close to a prescribed classical configuration $q_\text{cl}$. The filter function takes the form $\alpha_{\text{cl}}[q]=\int \mathcal{D}J e^{i\int dx J\pa{q(t)-q_{\text{cl}}}}\alpha_{\text{cl}}[J]$. We may write the decoherence functional between two classical histories in terms of the closed-path-time generating functional. In principle, we can examine adjacent general classical solutions for their consistency but, in practice, it is simpler to restrict ourselves to particular solutions $q_\pm$ according to the nature of the decoherence that we are studying.   Having considered all these issues, the decoherence functional results in
\begin{equation}
    \mathcal{D}[q^+_{\text{cl}},q^-_{\text{cl}}]\approx \mathcal{F}[q^+_{\text{cl}},q^-_{\text{cl}}],
\end{equation}
where $\mathcal{F}[q^+_{\text{cl}},q^-_{\text{cl}}] = e^{iS_{\text{IF}}[q^+_{\text{cl}},q^-_{\text{cl}}]}$ is the Feynman-Vernon influence functional defined by Eq. (\ref{funcional_influencia_definicion}).  Recalling the expression for the influence functional, once we have chosen the classical solutions of interest, adjacent histories become consistent at the time $t_{\text{D}}$ for which\newline $\Im S_{\text{IF}}|_{t_{\text{D}}}\approx 1$. As it has been noted, in practice the use of the decoherence functional 
looks to be less restrictive than the master equation, and suitable for problems in quantum field theory as well.

\subsection{Imaginary part of $S_{IF}^{(1)}$}
Let us recall the expression for $N_1(t,t')$ given by Eq. (\ref{noise}), and consider two classical trajectories $q_{\text{cl}}(t)$ wich differ in a phase factor
\begin{equation}
    \Delta q_{\text{cl}}(t)= q_0 \cos\pa{\omega_0 t+\delta}-q_0 \cos\pa{\omega_0 t}.
\end{equation}
We can write an expression for the imaginary part of the influence action considering the particle in presence of the vacuum field (ignoring the plate), which is
\begin{align}\nonumber
    &\Im \pa{S_{\text{IF}}^{(1)}}=-g^2 q_0^2\pa{1-\cos\pa{\delta}} \Im \int dt dt'\int \frac{dk}{(2\pi)^4} \\\nonumber
    & \pa{\frac{1}{k^2+i\epsilon}
-2\pi i \text{n}_B(|k^0|)\delta(k^2)}\left[e^{i\pa{k^0-vk^1+\omega_0}(t-t')} \right.\\\nonumber
&+e^{i\pa{k^0-vk^1-\omega_0}(t-t')}+e^{i\pa{k^0-vk^1+\omega_0}t}e^{i\pa{k^0-vk^1-\omega_0}t'}\\
&\left. + e^{i\pa{k^0-vk^1-\omega_0}t}e^{i\pa{k^0-vk^1+\omega_0}t'}\right].
\end{align}
Integration over the time variables result in Dirac delta functions. The last two terms vanish, as the conditions imposed by each of the multiplied deltas cannot be fulfilled simultaneously.
From each non-vanishing term, an infinite $\delta(0)$ is obtained, accounting for the total time of integration $T$ (time of flight of the particle). 
Taking then the limit $\epsilon \rightarrow 0$ we can use again the Sokhotski–Plemelj theorem \cite{weinberg}. We can also absorb the term with $\omega_0\rightarrow-\omega_0$ by the use of a change of variables $k^1\rightarrow-k^1$.
Applying properties of Dirac delta functions and changing $\{ k^2,k^3\}$ to cylindrical coordinates, we can perform the integral over $k=\sqrt{(k^2)^2+(k^3)^2}$ and $\theta_k$ to find
\begin{align}
      \Im & S_{\text{IF}}^{(1)}=\frac{g^2q_0^2T}{2}\pa{1-\cos\delta}
     \int \! \! dk^1\pa{2\text{n}_B(|k^1v-\omega_0|) \! + \! 1} \nonumber \\
     & \times \theta\pa{\pa{k^1v-\omega_0}^2-\pa{k^1}^2} \sqrt{\frac{\pa{k^1v-\omega_0}^2-\pa{k^1}^2}{\zeta\pa{k^1}}}  . \label{s1}
\end{align}
In the limit $v\ll 1$, Eq. \eqref{s1} reduces to
\begin{align}
     \Im S_{\text{IF}}^{(1)}\approx& \, \, \frac{g^2 q_0^2T}{4} \pa{1-\cos\pa{\delta}}\frac{\pi\w}{a} \\
     & \times \cor{ \pa{1+\frac{9v^2}{8}}\pa{2\nb{\w}+1}-f(\w\beta)v^2} \nonumber \, ,
    \label{s1_aprox}
\end{align}
where
\begin{equation*}
    f(\w\beta)=\w\beta \frac{\nb{\w}^2 e^{\w\beta}}{4} \pa{8+\w\beta\frac{e^{\w\beta}+1}{e^{\w\beta}-1}} \, .
\end{equation*}

\subsection{Imaginary part of $S_{IF}^{(2)}$}
Let us now recall the expression for $N_2(t,t')$ given by Eq. (\ref{noise}), and consider the same two classical trajectories $q(t)$. 
By inserting these results into Eq. (\ref{funcional_influencia_nucleos}), we are able to write an expression for the imaginary part of $S_{IF}^{(2)}$. Following the same procedure used to find $\Im S_{\rm IF}^{(1)}$ and the frictional force (\ref{fuerza}), we can express it as an integral over the dimensionless variables $\kk$ and $\y$ :
\begin{equation}
    \Im\pa{S_{IF}^{(2)}}=g^2 q_0^2T\frac{\lambda^2a^2}{4}\pa{1-\cos\delta}\cor{-\mathcal{S}_1+\mathcal{S}_2+\mathcal{S}_3},
\end{equation}
where the dimensionless terms $\mathcal{S}_i$ are
\begin{widetext}
\begin{align}\nonumber
    \mathcal{S}_1=& \int d\y  \frac{v}{2\oo} \theta \cor{\zeta_-\pa{k^2}}\left[\cos\pa{\frac{2}{v}\sqrt{\zeta_-\pa{k^2}}}\frac{2\nb{\oo}+1}{\zeta_-\pa{k^2}}+2\cos^2\pa{\frac{1}{v}\sqrt{\zeta_-\pa{k^2}}}\frac{8\nb{\oo}^3+8\nb{\oo}^2 +\nb{\oo}}{\vartheta}\right]\\[1.2em] \nonumber
    \mathcal{S}_2=&\int d\y d\kk \frac{1}{\pi}\frac{\sin\pa{2\sqrt{\zeta\pa{k^1,k^2}}}}{\sqrt{\zeta\pa{k^1,k^2}}}\text{p.v.}\pa{\frac{1}{(\kk v-\w)^2-\oo^2}}\left[\frac{1}{\sqrt{\zeta\pa{k^1,k^2}}}+\frac{2\nb{|\kk v-\w|}}{\sqrt{\zeta\pa{k^1}}}\right]\theta\cor{\zeta\pa{k^1,k^2}}\\[1.2em]
    \mathcal{S}_3 =& \int d\y \frac{v}{2\oo}(2\nb{\oo}+1)\left(\frac{e^{-\frac{2}{v}\sqrt{-\zeta_-\pa{k^2}}}}{\zeta_-\pa{k^2}}\theta \cor{-\zeta_-\pa{k^2}} - \frac{e^{-\frac{2}{v}\sqrt{-\zeta_+\pa{k^2}}}}{\zeta_+\pa{k^2}}\right),
    \label{s_1_2_3}
\end{align}
\end{widetext}

with $\zeta$ and $\vartheta$ defined as in Eq. (\ref{fuerza}). Again, the term $\mathcal{S}_1$ containing the theta function\newline $\theta \pa{\oo^2v^2-(\w-\oo)^2-\kk^2v^2}$ vanishes even for considerably high velocities of the particle. Then, for a non-relativistic motion, the imaginary part of $S_{IF}^{(2)}$ simplifies to
\begin{equation}
    \Im\pa{S_{IF}^{(2)}}=g^2 q_0^2T\frac{\lambda^2a^2}{4}\pa{1-\cos\pa{\delta}}\cor{\mathcal{S}_2+\mathcal{S}_3}.
\end{equation}
\subsection{Estimation of the decoherence time of the particle}

\begin{figure}[h]
\centering
\includegraphics[width=\columnwidth]{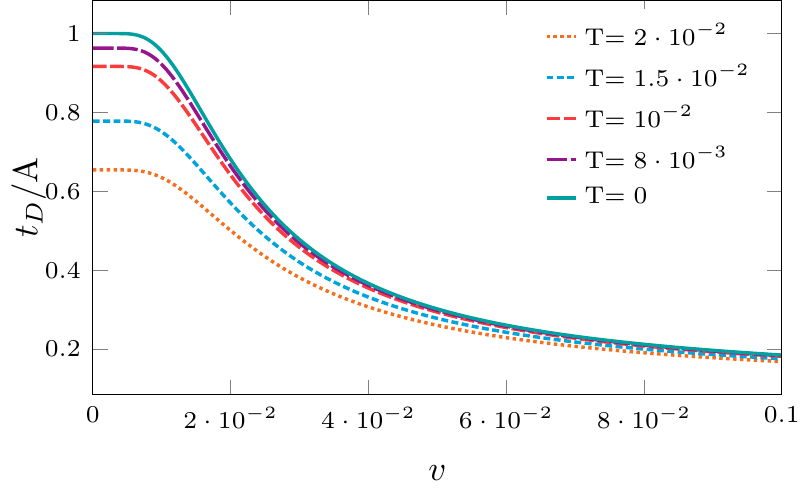}
\caption{\label{grafico_td_vs_v}Estimation of the decoherence time, in units of a global factor $\frac{4}{g^2 q_0^2\pa{1-\cos\pa{\delta}}}$ as a function of the relative velocity between the particle and the plate for fixed dimensionless frequencies $\w=0.03$ and $\oo=0.01$. }
\end{figure}

So far we have found a perturbative result for the imaginary part of the influence action (valid up to second order in $\lambda$ and non-relativistic velocities of the particle) given by  
\begin{align*}
    \Im\pa{S_{\text{IF}}}\sim\frac{g^2q_0^2T}{4}\pa{1-\cos\pa{\delta}} \Bigg\{\frac{\pi\w}{a} \Bigg[ \pa{1+\frac{9v^2}{8}}\\
    \pa{2\nb{\w}+1}-f(\w\beta)v^2] +\lambda^2a^2\cor{\mathcal{S}_2+\mathcal{S}_3}\Bigg\}.
\end{align*}
We are now able to estimate the decoherence time for the particle using the upper bound imposes by $\Im S_{\text{IF}}\sim 1$ . As we have detailed at the beginning of this section, the decoherence time can be estimated as the time of flight of the particle when this condition is satisfied, so
	\begin{align}
	t_{\text{D}}\sim & \frac{4}{g^2 q_0^2\pa{1-\cos\pa{\delta}}} \left\lbrace   \frac{\pi\w}{a} \pa{1+\frac{9v^2}{8}}\pa{2\nb{\w}+1} \right. \nonumber \\
	  & - \left.  \frac{\pi\w}{a} f(\w\beta)v^2 + \lambda^2a^2\cor{\mathcal{S}_2+\mathcal{S}_3} \right\rbrace ^{-1} \, ,
	\label{tiempo_de_decoherencia}
	\end{align}
where $S_1$ and $S_2$ are given by Eq. (\ref{s_1_2_3}).
After numerically performing the integrals appearing in Eq.(\ref{tiempo_de_decoherencia}), we present our results in Fig. \ref{grafico_td_vs_v} . Therein, we show the decoherence times estimated  made as a function of the relative velocity of the particle to the plate, for different temperatures and for a dimensionless coupling between the plate and the vacuum $a^{3/2}\lambda=0.01$. 

There is a global factor $A=\frac{4}{g^2 q_0^2\pa{1-\cos\pa{\delta}}}$ which shows that the decoherence time is reduced for larger values of the coupling constant between the particle and the vacuum field, and for certain values of the phase difference of the classical trajectories under consideration \cite{farias2016}. 
The behavior of the decoherence time plotted in Fig. \ref{grafico_td_vs_v}  shows that for any finite temperature (including zero-T) the particle's internal degrees of freedom suffers from decoherence due to the presence of the plate and the vacuum field.  As expected, the particle at rest ($v=0$) exhibits the already known behavior: smaller decoherence timescale  for higher temperatures; being the longest decoherence timescale for zero temperature \cite{PLA2004}. As the velocity of the particle is increased, the decoherence effects become more appreciable and less temperature-dependent, reaching a similar value for every temperature as $v$ tends to almost relativistic values. This is due to the fact we are estimating the decoherence time 
using an upper bound in the decoherence functional. This bound is satisfied quickly at high velocities at any temperature.
\begin{figure}[h]
\centering
\includegraphics[width=\columnwidth]{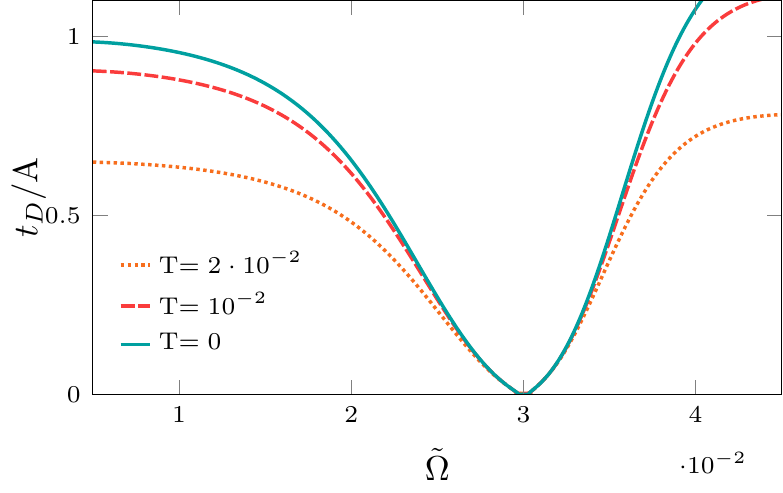}
\caption{Estimation of the decoherence time, in units of a global factor $\frac{4}{g^2 q_0^2\pa{1-\cos\pa{\delta}}}$ as a function of the dimensionless frequency of the harmonic oscillators forming the plate, for a fixed relative velocity $v=0.01$ and ${\tilde \omega}_0=0.03$. }
\label{grafico_td_vs_omega}
\end{figure}

In Fig. \ref{grafico_td_vs_omega}, the decoherence time $t_{\text{D}}$ is shown as a function of the plate's characteristic dimensionless frequency $\oo$, for different temperatures. For every temperature, the decoherence time tends to a minimum value at $\oo=\w$, meaning that the decoherence is maximal at the resonant case. The temperature dependence follows the same hierarchy shown in Fig. \ref{grafico_td_vs_v}.
From the results shown above we can see that, in this simple model, the decoherence time is reduced by temperature and velocity, being this reduction stronger in resonant situations.
 The decoherence effects could be maximized by an appropriate choice of the particle's fine-grained history (the phase difference $\delta$ in our example) and the plate's material (for a relation between the $\lambda(\omega)$ function of the material and its dielectric permittivity $\epsilon(\omega)$, see Refs. \cite{fosco_neumann,rubiolopez_2015}). Even for finite temperatures, we notice a velocity induced correction to the decoherence time, which could be used as an indirect way of detecting quantum friction.
It is worth mentioning the there is a subtle difference in the result obtained  if we had otherwise considered classical trajectories on phase but with different amplitudes instead of considering two classical trajectories differing in a phase (as some of us have done in our previous work \cite{farias_friction}). In such case, we would have obtained a result a similar   dependence on the parameters of the model as Eq. (\ref{tiempo_de_decoherencia}), but the factor $A$ would have been different, with $A=\frac{8}{g^2\Delta q_0^2}$.

\section{Conclusions}

The study of the friction force on an atom moving through the thermal radiation field in vacuum can be extensively found in the Literature, as for example in Ref.\cite{Mkrtchian}. It is usually  considered that the dissipation of the atom may be due either to the intrinsic radiation reaction of the electromagnetic field or to the dissipation in the metal surface in which the atom is reflected. Radiation reaction is an inescapable effect, but usually very small compared to dissipation within metals \cite{milton2016reviewfriction}. If the surface is metallic, the dissipation in the bulk is provided by the resistivity of the metal, or its conductivity. The dominant loss mechanism of the atom is due to the damping also provided by the surface; this can be thought of as the dissipation of the image of the atom moving through the bulk. Therefore, by using either the Kubo formalism or the fluctuation-dissipation theorem, it can be found  that the frictional force is proportional to 
$\alpha^2v^3/(\sigma a^{10})$  for low velocities, where $\alpha$ is the static polarizability of the atom (assumed isotropic), $v$ is its velocity parallel to the surface, $\sigma$ the conductivity, and $a$ 
is the distance between the atom and the surface \cite{Brevik2015}. The salient dependence is upon the cube of the velocity and the inverse tenth power of the distance. This friction force becomes appreciable only if the atom is extremely close to the surface. 
It is noteworthy that authors in Ref. \cite{intravaia_acceleration}  obtained a similar result as in Ref. \cite{barton_atom_halfspace},  by the use of a perturbative method.
Therein, authors state that the linear velocity dependence found in Ref. \cite{barton_atom_halfspace} is an artifact of the particular velocity profile assumed.  In Ref. \cite{Scheel} authors have also obtained such a linear dependence. 
 
However, it is important to emphasize that the existence literature on quantum friction force states different velocity dependence as the considerations on the material vary. Mainly, the frictional force critically depends on the dissipative mechanism assumed  in the material mirror. In the vacuum radiation field of empty space (without the presence of the plate), the atom would only suffer dissipation due to radiation reaction. The latter is due to emission and absorption of dipole radiation to obtain equilibrium with its surroundings. Due to this, the oscillations of the atomic polarization are damped leading to dissipation (altogether with the conductivity of the metal plate). If this happens to be the dominant mechanism, 
the frictional force on an atom traveling at constant $v$ at a distance $a$ of the plate yields a result proportional to $v^5$.  The change in power in velocity (with respect to the above scenario) is due to different microscopic (or macroscopic) models for the plate structure. 

Herein, we have studied, using the CTP approach, the corrections due to a finite temperature to the decoherence and dissipative effects suffered by a particle that moves above an imperfect mirror following a macroscopically prescribed trajectory, parallel to the plate and with constant velocity. We have explicitly calculated the frictional force, finding that it develops two components of different nature:  (i) a pure quantum component, present even at zero temperature and due exclusively to the existence of quantum vacuum fluctuations; and (ii)  a temperature-dependent component generated by the thermal fluctuations of the vacuum field that increases with the velocity becoming the main contribution at higher temperatures. We have also found that, for small but finite temperatures, both contributions are of the same order of magnitude. This would implied that the quantum frictional component could not be neglected.

It has been suggested that the thermal Casimir frictional force might be experimentally detected. This is due to the fact that thermal frictional force becomes significantly 
enhanced with respect to the zero temperature case. However, this conclusion might be unrealistic due to the extremely small separations $a$ considered. In this direction, we 
have used an approach that enable us to introduce thermal (and non-equilibrium) contributions in a microscopically-based model. In order to compute the frictional force we 
have developed the CTP formalism applied to the case of the particle coupled to the vacuum field and also to the microscopic degrees 
of freedom of the material. We have expanded up to four order, i.e. order two in the coupling between the particle and the field  times order two in the expansion in the 
coupling strength with the mirror degrees of freedom.  The CTP formalism is essential to obtain the correct result for $\langle 0_{\rm in} \vert t_{\mu\nu} \vert 0_{\rm  in} \rangle$ at finite temperature. The crucial point here is that, due to dissipation, the in and out vacuum states are different. This is the reason why the CTP formalism is not required to compute static Casimir forces, while its use is unavoidable to compute the force between the moving particle and the mirror \cite{fosco2007_moving_mirrors}. We have obtained analytical 
expressions for the force from our first principles approach.  The frictional force is shown in Fig. 2. From an expansion in powers of the parallel velocity $v$, we can see that 
our result predicts a linear behavior of the force. This is in accordance with some of the references above, and we believe that it is mainly due to the model used to describe the 
microscopic structure of the mirror. In this sense, the absence of absorption in the material could be responsible for this result. We expect that a more general model, considering 
the mirror composed by a dissipative but also absorbing material, could give more conclusive answers about this aspect of quantum friction.

Finally, we have also calculated the decoherence time, and found its dependence upon both velocity and temperature. We have found that while decoherence effects are enhanced as both quantities are increased, the effect-dependence upon the temperature is more relevant for smaller velocities. This contrasts the result obtained for the frictional force, which is highly dependent on the temperature for greater velocities. This result implies that, for higher velocities, a reduction in the decoherence time at any temperature would be mainly the result of quantum vacuum fluctuations, and thus the detection of such reduction would be an indirect evidence of the existence of quantum friction. This fact was presented, firstly, in Ref.\cite{farias2016}, and it is now complemented with the thermal corrections that appear as relevant in determination of the decoherence effect over the internal degree of freedom of the 
moving atom.  We expect that, if decoherence due to these frictional effects is enhanced, more viable is the possible scheme to measure quantum friction indirectly measuring decoherence effects.

The approach described for the calculation of the frictional force (and hence for decoherence effects) can be generalized to the more realistic case of the electromagnetic field at non-vanishing temperature. The nonlocal interaction should be generalized accordingly and will involve the derivatives of the potential vector $A_\mu$ on the position of the mirrors. Also the inclusion of absorption into the material model may be added. Work in this direction is in progress.
\newline
\newline
\section*{Acknowledgements}

This work was supported by ANPCyT, CONICET, and Universidad de Buenos Aires; Argentina. 

\bibliographystyle{apsrev4-1}
\bibliography{friction.bib}




\end{document}